\documentclass[11pt,a4paper]{emulateapj}
\usepackage{epstopdf}
\usepackage{amsmath}
\usepackage{natbib}

\newcommand{\Msun}{{M_{\odot}}}

\newcommand{\HI}{H{\sc ~i}}

\newcommand{\HeI}{He{\sc ~i}}
\newcommand{\HeII}{He{\sc ~ii}}

\newcommand{\OVI}{O{\sc ~vi}}

\usepackage{color}

\begin{document}

\title{The Sources of Extreme Ultraviolet and Soft X-ray Backgrounds}
\author{Phoebe R. Upton Sanderbeck\altaffilmark{1}, Matthew McQuinn \altaffilmark{1},  Anson D'Aloisio\altaffilmark{1,2},  Jessica K. Werk\altaffilmark{1}}
\altaffiltext{1} {Department of Astronomy, University of Washington\\}
\altaffiltext{2} {Department of Physics and Astronomy, University of California, Riverside\\}

%
%

\begin{abstract}
 Radiation in the extreme ultraviolet (EUV) and soft X-ray holds clues to the location of the missing baryons, the energetics in stellar feedback processes, and the cosmic enrichment history.  Additionally, EUV and soft X-ray photons help determine the ionization state of most intergalactic and circumgalactic metals, shaping the rate at which cosmic gas cools.  Unfortunately, this band is extremely difficult to probe observationally due to absorption from the Galaxy. In this paper, we model the contributions of various sources to the cosmic EUV and soft X-ray backgrounds.   We bracket the contribution from (1) quasars, (2) X-ray binaries, (3) hot interstellar gas, (4) circumgalactic gas, (5) virialized gas, and (6) super-soft sources, developing models that extrapolate into these bands using both empirical and theoretical inputs.  While quasars are traditionally assumed to dominate these backgrounds, we discuss the substantial uncertainty in their contribution.  Furthermore, we find that hot intrahalo gases likely emit an ${\cal O}(1)$ fraction of this radiation at low redshifts, and that interstellar and circumgalactic emission potentially contribute tens of percent to these backgrounds at all redshifts.  We estimate that uncertainties in the angular-averaged background intensity impact the ionization corrections for common circumgalactic and intergalactic metal absorption lines by $\approx 0.3-1~$dex, and we show that local emissions are comparable to the cosmic background only at $r_{\rm prox} = 10-100~$kpc from Milky Way-like galaxies.

\end{abstract}

\section{Introduction}
\label{sec:intro}

The extreme ultraviolet (EUV) through soft X-ray represent a slice of the electromagnetic spectrum that is difficult to observe in astronomical spectra.  Not only must it be observed from above Earth's atmosphere, but for extragalactic sources much of this slice is absorbed by the Galaxy.  Yet, because most transitions to ground Rydberg states of ions fall in this band and, hence, most cooling emissions, the EUV/soft X-ray extragalactic background holds clues into the location of the missing baryons, the energetics in stellar feedback processes, and the cosmic enrichment history.  The lack of observations in this band complicates inferences from observable ionic transitions because these backgrounds often shape the ionic ratios of diffuse astrophysical gases.

AGN, or more specifically the brightest of these, quasars, are thought to be the dominate extragalactic source for radiation in the EUV and soft X-ray, especially at $z\lesssim 3$ \citep{meiksin93, madau99, faucher08, haardt12}.\footnote{Starlight is an important contribution at $1-4$Ry in some models (and dominant at lower energies), with its relative importance depending on how efficiently these photons can escape from galaxies.}  However, quasars' spectral energy distributions (SEDs) have only been measured at frequencies redward of $\sim 25$ eV \cite[e.g.][]{telfer02}, and current theoretical models are likely not successful enough to motivate extrapolations to the soft X-ray.   Thus, models for quasar spectra extrapolate with a single power-law between the EUV and soft X-ray, even though the quasar SED is certainly more complex \citep{bechtold87, laor97,done12}.

Quasars are not the only source that may contribute substantially to this critical waveband: X-ray binaries (XRBs), cooling gases (10$^4$-10$^7$ K) in the interstellar medium of galaxies (ISM) and the circumgalactic medium (CGM) that surrounds them, and virialized halo gas may also substantially contribute. Indeed, XRBs tend to dominate galactic X-ray emissions at $\gtrsim 1$ keV, while at lower energies radiative cooling from the hot ISM in galaxies is thought to become important \citep{grimes05, tullmann06, owen09, mineo12b, lehmer15}.  This hot gas is the result of feedback from supernovae, massive star winds, and radiation pressure \citep{1977ApJ...218..148M, chevalier85, strickland00}.  A significant fraction of this feedback could also be radiated in the surrounding CGM in addition to the hot ISM \citep[e.g.,][]{mcquinn17}. 
  Similarly, intrahalo gas that has been shocked by virialization can also be important, especially with decreasing redshift as more massive and, hence, hotter systems form.  

Many fields of study are reliant on models for the metagalactic ionizing background.  Such models are used in essentially all calculations of the ionization state of photoionized gas in the intergalactic medium (IGM) and CGM \citep[for a recent review see][]{2016ARA&A..54..313M,tumlinson17}, and they are used to set the ionization state of gas in cosmological simulations (and, hence, shape the rate at which simulated gas cools).  However, previous ionizing background models only considered the contribution of quasars in the EUV and soft X-ray, making the typical simplifying assumptions to extrapolate from observed bands \citep{haardt96, faucher09, haardt12}.  Even if the background is dominated by quasars, these models may not apply near galaxies (where they are most used), since many UV and X-ray sources are (circum)galactic in nature \citep{2010MNRAS.403L..16C, 2012ApJS..202...13G}.  

This paper estimates the contribution of all source classes that we are aware of that may emit substantially in the EUV and soft X-ray, in the process estimating the plausible range of $z\sim0$ and $z\sim 2$ ultraviolet background angular-averaged specific intensities, $J_\nu$.    Section~\ref{sec:sources} estimates the emissivities of various EUV and soft X-ray sources.  Using these estimates, we model the relative contribution of these sources to the ionizing background in Section~\ref{sec:jnu}. Section~\ref{sec:cgm} discusses the impact of the uncertainties in the background on the ionization of highly ionized metals. Finally, for galactic sources, we estimate the size of the local enhancement expected around each galaxy in Section~\ref{sec:prox}.

\section{Modeling the sources}
\label{sec:sources}
This section describes all the known sources that emit substantially in the UV and soft X-ray, developing models for their specific emissivity across redshift.  These calculations are then fed into calculations of the background $J_\nu$ in Section~\ref{sec:jnu}.

\subsection{Active galactic nuclei (AGN) }
\label{sec:QSO}

Quasars -- highly luminous AGN -- are likely the most emissive sources in the EUV and soft X-ray.  Yet, their emissivity in this band is highly uncertain. Ultraviolet observations of quasar emissions can only be used to reliably estimate the mean SED to wavelengths as short as several hundred Angstroms \cite[e.g.][]{telfer02}.    A break is observed in the mean SED at $\lambda \sim1000\;$\AA, with the spectrum softening blueward of this break.  This break is commonly attributed to the innermost temperature of the quasar's accretion disk (`the Big Blue Bump'; \citealt{1973A&A....24..337S}) and the transition of quasar spectra to Compton up-scattered radiation from a coronae \citep[e.g.][]{czerny87}.   However, thin accretion disk plus corona models have limited success at describing quasar spectra in the UV \citep{laor97, davis07}, which may be partly cured by invoking inhomogeneous temperatures \citep{2011ApJ...727L..24D} and opacity effects \citep{czerny87}.   Additional tuning must be added to achieve power law-like emission into the soft X-ray as is observed \citep{bechtold87, laor97}:  The disk emission almost certainly needs to be Compton up-scattered by material distinct from the traditional coronae. This could be accomplished by an optically thick region around the disk interior at temperatures of $\sim0.2~$keV \citep{done12}.

\begin{figure}
\begin{center}
\resizebox{9.0cm}{!}{\includegraphics{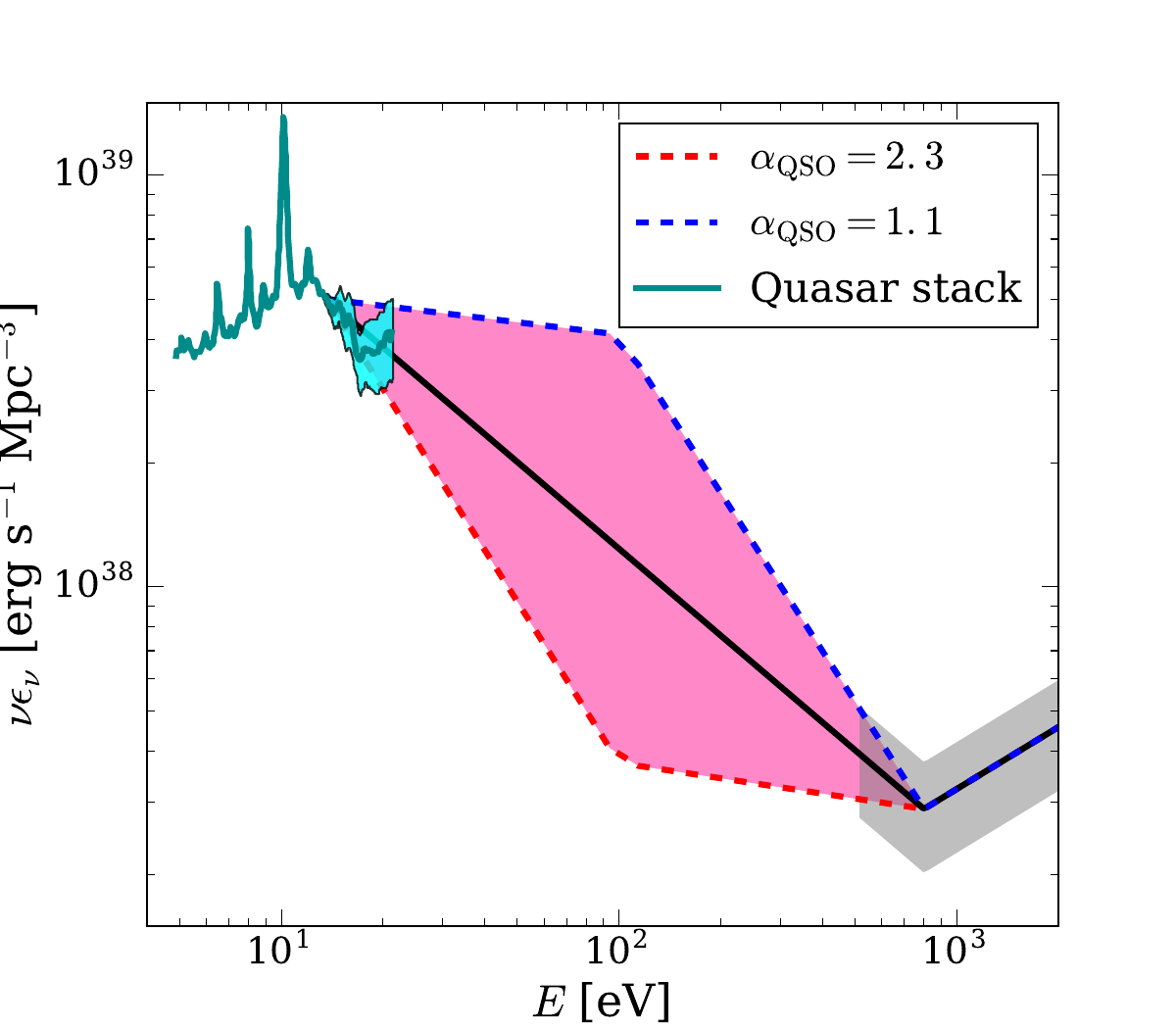}}\\
\end{center}
\caption{Observations and stacks of the quasar specific emissivities. The jagged turquoise solid curve shows the stacked quasar SED of \citet{lusso15}, normalized to match the \citet{khaire15} specific emissivity at 1~Ry. Filled regions represent the range of allowed values. The solid line at $13.6-800$~eV is the fiducial power-law model with $\alpha_{\rm QSO} = 1.7$. The  \citet{lusso15} 1$\sigma$ upper and lower limits on $\alpha_{\rm QSO}$ are shown by the blue and red dashed lines.  The inner pink shaded region represents the range of viable specific, which are tapered to $800$eV to match soft X-ray observations.  The harder component above $800$eV owes to coronal emission.
\label{fig:qso}}
\end{figure}

Because of these open issues, UV background models have used, rather than theoretically-motivated templates, simple power-laws estimated from quasar stacks to extrapolate to shorter wavelengths.  These power-laws fit stacked spectra blueward of the break at $\sim1000~$\AA\ and redward of the highest energy where the SED can be reliably estimated ($\sim500$\AA).   There is limited dynamic range over which the power-law is fit. The \citet{lusso15} stacked SED is the solid turquoise curve in Figure~\ref{fig:qso}, with the  cyan region highlighting the estimated 1-$\sigma$ errors.  We have normalized the SED to the $z=0$ specific emissivity at $1~$Ry as described below. 

The effective power-law index of the quasar stacks, $\alpha_{\rm QSO}$ (see Equation~\ref{eq:pow}), that analyses estimate varies significantly, with \citet{telfer02} finding $\alpha_{\rm QSO} = 1.57\pm 0.17$ for their radio quiet sample, \citet{shull12} and \citet{stevens14} finding $\alpha_{\rm QSO} = 1.4\pm 0.2$, and \citet{lusso15} finding $\alpha_{\rm QSO} = 1.7\pm 0.6$, with their larger error owing to analysis differences described below.  These observations show rough consistency (although they are discrepant with the lower luminosity sample of \citet{scott04}, which finds $\alpha_{\rm QSO} = -0.6\pm 0.4$).  Lastly, photoionization calculations for the broad UV lines in these spectra favor somewhat softer spectra with $\alpha_{\rm QSO} \approx 2$ \citep{baskin14, lusso15}.

\citet{lusso15} pointed to several factors that result in differences in the estimated $\alpha_{\rm QSO}$ from stacking.  Especially for quasars at $z\geq1$, estimates for the stacked SED must correct for \HI\ absorption from the IGM (with the Lyman series affecting $\lambda < 1216~$\AA\ and Lyman-continuum $\lambda < 912~$\AA). The differences in how this correction is done lead to some of the discrepancies in the reported values and their uncertainties.  In addition, where the $\sim 1000$\AA\ break wavelength is placed affects the final slope, and many samples do not have much spectral coverage of this break. Finally, broad line features are present in the stacks and so different choices are made for how to fit the continuum underneath the lines.

A justification often given for extrapolating with a single power-law into the soft X-ray is that the spectral indices of Type 1 AGN required in the EUV is similar to the spectral index needed to extrapolate to soft X-ray band, as well as the power-law index found in the soft X-ray \citep{laor97}.  However, both in the UV and soft X-ray, individual systems show broad dispersion in their allowed index (although individual system inferences are difficult in the UV owing to IGM absorption).  In addition, within the range of allowed mean spectral indices, there is no reason why there cannot be a softening followed by a hardening (indeed a soft spectrum followed by a hotter coronae spectra is seen in models such as \citealt{done12}) or vice versa (in the stacked SED, some quasars should harder and, hence, become more dominant at bluer wavelengths, hardening the spectrum).  

Figure~\ref{fig:qso} shows the empirical estimates and extrapolation of quasar emissivity at $z\sim 0$. For our calculation, we use quasar emissivities reported at $1~$Ry from \citet{khaire15} that is consistent with the mean Ly$\alpha$ forest transmission \citep{fumagalli17}. These emissivities are extrapolated from the g-band luminosity function with the relation $\log{\epsilon_g} = \log{\epsilon_{912}} + 0.487$ \citep{khaire15,haardt12,boyle00,croom04}, and since the mean SED is much better known between these wavelengths -- and the normalization is essentially calibrated to match robust constraints at $1~$Ry from the Ly$\alpha$ forest -- this extrapolation should not contribute significantly uncertainty. 

We extrapolate from this emissivity at $1$Ry with a power-law spectrum parametrized as
\begin{equation}
J(\nu)=J_0\left(\frac{\nu}{\nu_{1{\rm Ry}}}\right)^{-\alpha_{\rm QSO}}.
\label{eq:pow}
\end{equation}
We vary $\alpha_{\rm QSO}$ between $1.1$ and $2.3$, the $1\sigma$ error bar of \citet{lusso15}, a range shown by the dashed curves in Figure~\ref{fig:qso}.  However, one cannot extrapolate all the way to $\approx 1~$keV with this spectrum unless it has $\alpha_{\rm QSO}\approx 1.7$ \citep{laor97} to be compatible with soft X-ray background measurements discussed in \S~\ref{sec:jnu}.  We taper our spectrum starting at $100~$eV so that it converges to the same point at $800~$eV as if $\alpha_{\rm QSO}=1.7$.  While our choices here are somewhat arbitrary, we believe it provides a reasonable guess for the allowed range of quasar specific emissivities (motivated both by the error on UV extrapolations and the large theoretical uncertainty in the spectral form).

Above $800~$eV, we assume the spectrum scales with a spectral index of $\alpha=0.5$.  While $800~$eV is on the lower side of the observations for where the harder coronal emission dominates, finding $1-2~$keV \citep{laor97}, this spectral index is characteristic of the coronae of radio loud quasars, and somewhat harder than that found for radio quiet ones (although \HI\ absorption additionally acts to harden their spectrum; \citealt{laor97}).  The selected slope of $\alpha=0.5$ is a good fit to the slope found in the more detailed models for this component in \citet{haardt12} and also to $z=0$ soft X-ray background measurements.  However, our tapering to a single curve at $>800~$eV, which is done to match the background intensity measurements in the soft X-ray, should underestimate the uncertainty in $\epsilon_\nu(z)$ at these energies.  We justify not attempting to quantify the uncertainty $>800~$eV because these higher energies are less important for this study.

\subsection{X-ray binaries (XRBs)}
\label{sec:XRB}

 The X-ray emission from galaxies is thought to be dominated by XRBs at $>1~$keV. The galactic XRB emission from high mass XRBs should trace the star formation rate -- as their formation lags star formation episodes by just tens of megayears -- and low mass XRBs should trace stellar mass -- as theirs can lag star formation by many gigayears.  These correlations have been confirmed observationally \citep{ranalli03, grimm03, gilfanov04, hornschemeier05, lehmer10, boroson11, mineo12a, zhang12}. We use these correlations to relate the cosmic XRB emissivity with redshift to the observed star formation rate density and stellar mass density.   We approach the XRB contribution to the EUV/soft X-ray from both a theoretical and an empirical standpoint, considering two models:

\paragraph{Empirical Model} In this model, all X-ray emission traces either the star formation rate or the stellar mass \citep{lehmer10}. Low-mass XRBs are likely as important as their high-mass counterpart in the average galaxy at $z\sim 0 $, though at $z\gtrsim 1 $ and in star forming galaxies, high-mass XRBs dominate the X-ray emission from XRBs. We allow the relationship between star formation rate and X-ray emission to account for temporal trends that result from such things as increasing cosmic metallicity.\footnote{Increasing metallicity is expected to decrease the number of XRBs because metallicity increases mass loss from stellar winds, affecting binary separations, and because higher metallicity stars have larger radii, which increases Roche lobe overflow \citep{linden10}. There is also observational evidence for X-ray flux scaling inversely with metallicity \citep{brorby16}.}  For high-mass XRBs (HMXBs), we use the redshift-dependent relation constrained in \citet{dijkstra12}, parametrized as $L_{X,0.5-8}/{\rm SFR} = [L_{X,0.5-8}/{\rm SFR} ]_{z=0} (1+z)^b$, where $L_{X,0.5-8}$ is the $0.5-8$~keV X-ray luminosity.    \citet{dijkstra12} finds $0\leq b \leq1.3$, anchoring to the \citet{mineo12a} best fit of $[L_{X,0.5-8}/{\rm SFR} ]_{z=0} =2.6\times10^{39}$ erg s$^{-1}$ (M$_{\odot}$ yr$^{-1})^{-1}$.   For the cosmic star formation rate density, we use the fit in \citet{haardt12} derived from rest-frame optical and UV observations.\footnote{The total SFR density that is often estimated is likely less appropriate because star formation enshrouded behind gas and dust is unlikely to produce EUV and soft X-ray photons that can escape.} For low-mass XRBs (LMXBs), we use a correlation between total stellar mass and X-ray luminosity, $L_{X,0.5-8}/{M_{\star}}\sim 9\times10^{28}$ erg s$^{-1}$ \citep{gilfanov04,lehmer10}. We use \citet{wilkins08} for the cosmic stellar mass density as a function of redshift.

This model's second ingredient is the XRB SED.  Above $300$ eV the SED of galactic X-ray emission has been constrained observationally.  For the SED, we stack the spectrum of four star-forming galaxy SEDs observed from $0.3-3~$keV with \textit{Chandra/XMM-Newton} \citep{lehmer15,wik14,yukita16}.  The stack is computed by averaging the specific luminosity per unit star formation rate (SFR) of each galaxy, $L_X(\nu)/$SFR.    However, the spectrum of XRBs is not constrained below $300$~eV.  To extrapolate, we model the XRB spectrum as a simple power law in specific luminosity with $\alpha_{\rm empirical} = 0.7$ (using the same conventions as with $\alpha_{\rm QSO}$ in eqn.~\ref{eq:pow}).  This slope is based on the scaling of stacked X-ray emission -- not just the four we use for our empirical spectrum \citep{rephaeli95,swartz04}.  

\begin{figure}
\begin{center}
\resizebox{9.0cm}{!}{\includegraphics{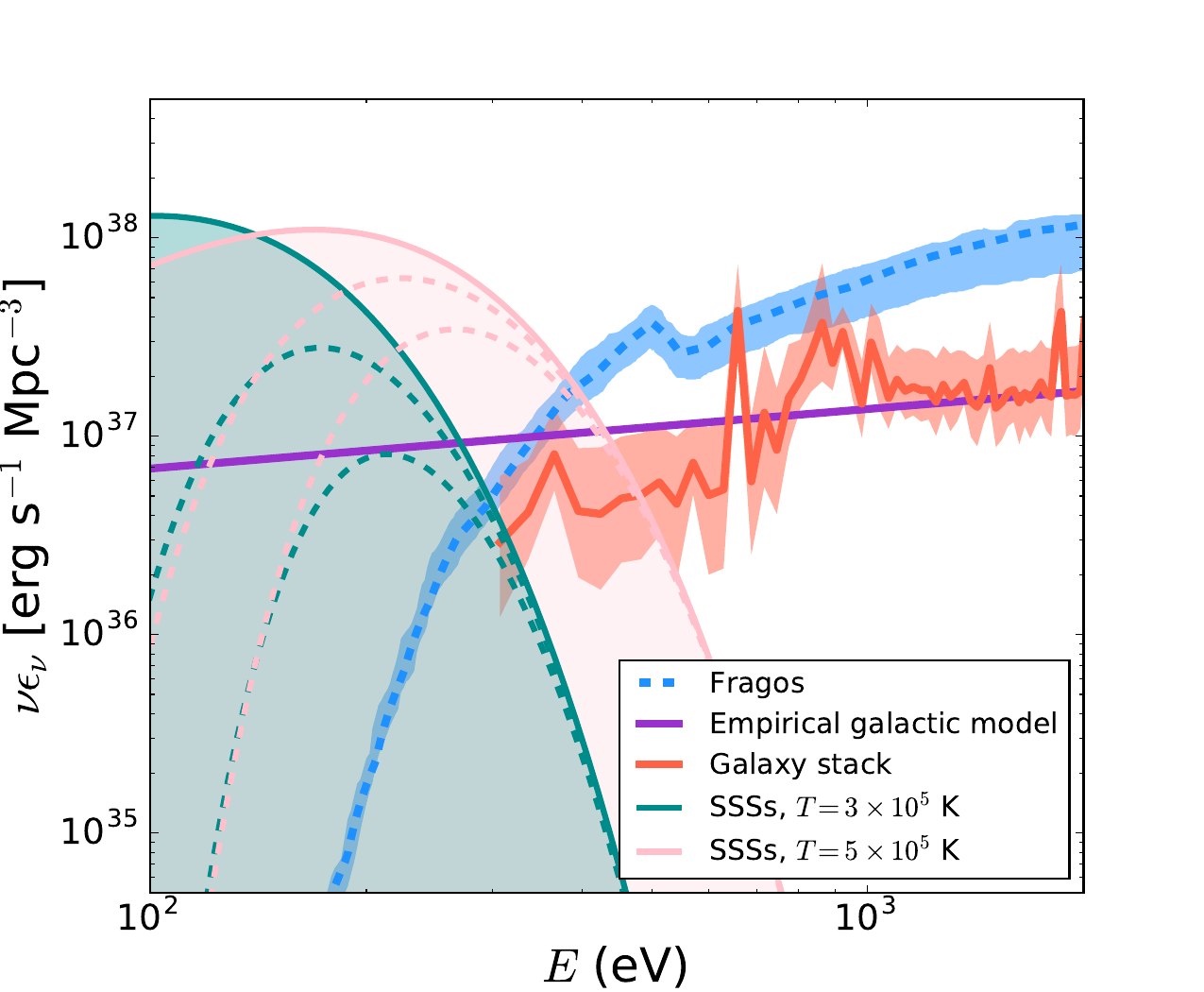}}\\
\end{center}
\caption{  EUV and soft X-ray models and bounds for the $z=0$ specific emissivities of XRBs and SSSs. The solid purple and red curves represent respectively the empirical models described in the text assuming the power-law SED and the stacked four star-forming galaxies.  The blue dashed curve represents the theoretical model of \citet{fragos13b}.  The highlighted bands around the red solid and blue dashed curves are, respectively, the standard deviation estimated from the four stacked galaxies and the quoted uncertainties in \citet{fragos13b}. The teal and pink shaded areas represent the allowed emissivity for SSSs with blackbody temperatures of $3\times 10^5$~K and $5\times 10^5$~K respectively, using the luminosity bounds from \citet{chen15}. The respective dashed curves represent this emission with attenuation from a $10^{20}$ cm$^{-2}$ and $3\times10^{20}$ cm$^{-2}$ column.}
\label{fig:XRB}
\end{figure}

The thick solid curves in Figure~\ref{fig:XRB} (and later Fig.~\ref{fig:ISM_eps}) show the resulting specific emissivity at $z=0$ for the stacked SED and $\alpha_{\rm empirical} = 0.7$ models.  Even among the four stacked galaxy spectra there is $\sim 1~$dex scatter in $L_X(\nu)/$SFR, owing largely to scatter in the number of ultra-luminous X-ray sources.  An extremely crude estimate of the error on the mean stack is shown by the highlighted regions in Figure~\ref{fig:XRB}, which is the standard deviation of the four spectra divided by $\sqrt{N_{\rm gal}-1}$.
 
 We note that much of the emission below $1~$keV is likely from hot gas in the ISM and not XRBs.  Most XRB's emission is found to decline below 1 keV rather than have a flat power-law as our extrapolation assumes (because of their intrinsic spectrum and \HI\ absorption), and  the $<1~$keV appears to owe to more diffuse sources as discussed in Section~\ref{sec:ISM}.  Thus, our $\alpha_{\rm empirical} = 0.7$ model may be better conceived as an amalgam of XRB and hot ISM emission, where hot ISM takes over at lower energies and has a power law-like spectrum.  We discuss the realism of this power-law extrapolation for ISM emission in Section~\ref{sec:ISM}. In addition, the bounds we place on super soft sources in \S~\ref{ss:SSS} also are relevant to potential soft emissions from LMXBs.

\paragraph{Fragos++(2013) Model  }   We use the XRB population synthesis model of \citet{fragos13b}.  It uses the XRB population synthesis simulations \texttt{StarTrack} \citep{belczynski08}, which are calibrated to observations of XRBs at low redshift, and the semi-analytic galaxy catalog of \citet{guo11}, which models the star formation and stellar metallicity histories.  The assumed SED is based on samples of observations of both neutron star and black hole XRBs.  The overall shape of their XRB SED does not change significantly with redshift, despite accounting for trends in metallicity and the fraction of LMXBs to HMXBs. Emissivities at high redshift will be slightly enhanced relative to star formation rate density due to the lower metallicity of the stars in binary systems. This reduces stellar winds and increases Roche lobe overflow, increasing the number of XRBs, especially HMXBs. The spectrum falls off rapidly at energies lower than a few hundred electron volts, due both to the XRB SED and interstellar/circumstellar absorption associated with most XRBs. The dashed blue curve in Figure~\ref{fig:XRB} shows this model, with the fainter blue band signifying their estimated uncertainties.\\

\subsection{Super soft X-ray sources}
\label{ss:SSS}
Super soft X-ray sources (SSSs) are observed systems with blackbody spectra that peak in the soft X-ray and that are theorized to be compact binary systems with a white dwarf accreting from either a main sequence star or a red giant and this accretion igniting on the surface \citep{kahabka97, greiner00}.  Accretion onto white dwarfs should be generic and such accretion is necessary in a major progenitor scenario of Type~1a supernovae.  Models find that the accretion should be thermally stable with rates of $\dot M\sim10^{-7}M_{\odot}$~yr$^{-1}$  \citep{nomoto07}. With such $\dot M$, accreting white dwafs should emit substantially in the $0.1-0.7$ keV band.   SSS could increase ISM photoheating rates enough to explain the higher temperatures inferred from photoionized ISM observations compared to models \citep{woods13}.

However, the number of SSSs that have been detected is 1-2 orders of magnitude below expectations from population synthesis models \citep{chen14,chen15}.  The lack of sources could be because of obscuring \HI\ columns around these sources (although see \citealt{nielsen15,woods16}), if dynamically unstable mass loss for giant stars occurs differently from in current models \citep{chen15}, or if the nature of accretion onto these systems is different \citep{cassisi98}. Observations are sensitive to SSSs with a more massive white dwarf primary, so the smaller primaries (which are more important at lower energies) could still be present at forecast abundances.  

Here we assume that stable accretion onto undetected (i.e. less massive) white dwarfs is occurring, and we attempt to place an upper bound on the total ionizing emissions.  We assume that the spectrum of these objects follows a blackbody with an effective temperature ranging between $3\times10^5$ and $5\times 10^5$~K, characteristic of the lower mass primaries that would go undetected in previous soft X-ray searches \citep{greiner00}.

For long-term thermally stable accretion onto a white dwarf in the stable hydrogen-burning regime, the ideal donor star is a 2M$_{\odot}$ main sequence star  \citep{chen14}. It follows that the luminosity of SSSs is expected to peak at stellar ages of approximately 1Gyr (the lifetime of a 2M$_{\odot}$ main sequence star) and drop off steeply at ages $>1$Gyr in the soft X-ray\footnote{4 Ry emission drops off less steeply, but still peaks at 1Gyr} \citep{chen15}, thus, globally, their emissions should peak at roughly $z\sim 1.8$, approximately 1Gyr following peak cosmic star formation rate.

To estimate the contribution of SSSs, we constructed the shape of the emissivity function from the shape of cosmic star formation rate density and shifted this evolution by 1Gyr.\footnote{Describing this evolution with the more complex lightcurve in \citet{chen15} should not change results over this 1Gyr-delayed model.} We normalize the upper bound to the emission from SSSs by setting the integrated $\geq 4$Ry luminosity from SSSs to equal the HeII-ionizing luminosity of the a025qc15 model in \citet{chen15} (which used binary Population synthesis models to predict the stellar age-specific luminosity). In this model, the criterion for stable mass loss for binaries with giant donors (a critical mass ratio) is adjusted to better agree with late-time soft X-ray flux of elliptical galaxies, which is over predicted with the more standard prescription. Their other two models that adjust the critical mass criterion to satisfy the X-ray limits increase this critical mass ratio, and thus show lower luminosities. However, their model without this prescription overpredicts the observational constraints in the X-ray. We normalize to the He-ionizing model rather than normalizing to the soft X-ray predictions because the soft X-ray energies fall in the tail of the blackbody and thus the normalization would be sensitive to the assumed spectrum. 

 Figure~\ref{fig:XRB} shows the allowed emissivity of SSSs, where teal and pink shaded areas represent the allowed emissivity for SSSs with blackbody temperatures of $3\times 10^5$ K and $5\times 10^5$ K respectively.  Although unlike \citet{chen15}, we use a single temperature for our spectral model, the limited range of temperatures consistent with stably accreting white dwarfs and allowed by observational constraints in the soft X-ray justifies this approximation. The dashed curves show the attenuation of this emission from a column of $10^{20}$ cm$^{-2}$ and $3\times10^{20}$ cm$^{-2}$. These models show upper bounds skimming the lower bounds of quasars at $\sim100$~eV. Our bound on SSSs allows a potential contribution that is as much as an order of magnitude above our empirical XRB model, barring any absorption. 

Accreting white dwarfs seem to be a natural consequence of stellar evolution -- it should then follow that SSSs contribute substantially to the ionizing background. However, there is limited observational evidence to support their ubiquity. If SSSs are numerous, their observational elusiveness requires a limited mass range of accreting WDs and gas-poor environment \citep{woods16} to respectively explain the lack of detections in the soft X-ray and the lack of ionized nebulae associated with these objects (although the lack of ionized nebulae may suggest that the bulk of their ionizing photons escape their host galaxy).

\subsection{Warm--Hot gas in the ISM and CGM}
\label{sec:ISM}
The ISM is a mix of different gas phases, with this multiphase structure driven largely by stellar wind bubbles and supernovae blastwaves \citep{1977ApJ...218..148M}.  The spectrum of radiation emitted by these bubbles and blastwaves depends on the resulting temperatures and ionization states of associated gases.  Some of this emission escapes into intergalactic space, with the amount depending on the \HI\ columns out of the host galaxy. Furthermore, some of the gas driven by these feedback processes vents from galaxies, potentially heating the CGM and leading to additional cooling emission that has less difficulty escaping from the galactic environs.

Radiation from warm-hot gas in the ISM can in principle be observed directly at $>300~$eV.  However, there is no robust determination of this ISM X-ray emissivity because no straightforward method exists for disentangling the ISM contribution from the XRB one.  \citet{mineo12b} attempted to disentangle the ISM contribution by subtracting off flux from X-ray point sources, which are likely XRBs, finding that most of the observed flux remains at the lowest energies observed, with little above $1~$keV. Using the remaining flux as an estimate for the ISM contribution, they measured the relation between ISM emission and SFR:
\begin{equation}
L_{X, 0.5-2}/{\rm SFR}\approx 8.3\times10^{38} {\rm ~ erg ~s^{-1} } ~(M_{\odot}~{\rm yr}^{-1})^{-1},
\label{eqn:mineo}
\end{equation}
where $L_{X, 0.5-2}$ is the $0.5 -2$ keV ISM luminosity.  

Observationally the shape of the average ISM SED is essentially unconstrained at $<300$eV, and the more diffuse emission from the CGM is unconstrained at all wavelengths. To model the spectral shape, we use the spectrum of gas cooling from some maximum temperature to a much lower temperature.  The specific emissivity of a population of objects with this spectrum is then given by
\begin{equation}
\epsilon_\nu(T_{\rm max}) = \dot \rho_{\rm SFR}  \times [L_{\rm bol}/{\rm SFR}] \times  \int_{T_{\rm min}}^{T_{\rm max}}  dT ~\epsilon_{T,\rm bol}^{-1} ~ \epsilon_T(\nu), 
\label{eqn:enu}
\end{equation}
where $\epsilon_{T,\rm bol} = \int d \nu  \epsilon_T(\nu)$, $ \dot \rho_{\rm SFR}$ is the UV star formation rate, and $[L_{\rm bol}/{\rm SFR}]$ sets the normalization (which we develop models for shortly).  We compute $ \epsilon_T(\nu)$ using {\sc Cloudy} \citep{ferland13} and assuming collisional ionization equilibrium\footnote{To understand the effect of ionizing backgrounds on our predictions, we have computed the ionization state with the Haardt and Madau 1996 model and $nT = 10$K~cm$^{-3}$ and  $nT = 100$K~cm$^{-3}$, thermal pressures justified in \citet{mcquinn17}.  We find there is a negligible difference in the spectra and thus the photoionization background does not have a significant effect on the resulting SED of the hot gas.}. We choose $T_{\rm min} = 10^4$K, although our calculations are not sensitive to this choice.   In addition, our calculations assume a single metallicity of $Z=0.3 Z_\odot$ for $T_{\rm max} =10^6$K and $Z=Z_\odot$ for $T_{\rm max} =3\times10^6$K. However, since metal emission lines dominate the bolometric emission -- what we normalize to -- metallicity has little effect on the resulting spectrum.

We choose $T_{\rm max}$ to have values of $1\times 10^6~$K or $3\times10^6~$K. The motivation for this spectral form is that gas tends to be shock heated to these temperatures by $10^2-10^3~$km~s$^{-1}$ flows before cooling.  At these temperatures, cooling times tend to be short and so there is not often a heating process to halt cooling from running away.   (Since we assume a single maximum temperature, our calculation misses that decelerating blastwaves tend to shock gases to a range of temperatures.)  In the CGM, our simple model has the additional motivation that gas may be cooling from the $\sim 10^6$K virialized phase characteristic of star forming galaxies. 

This model assumes CIE cooling, which is likely not a good approximation for ISM cooling. \citet{mcquinn-Xray} used \citet{allen} models for unmagnetized supernovae shocks to compute the spectrum of a supernovae blastwave as it decelerates in a constant density medium of density $1~$cm$^{-3}$. This calculation includes the non-equilibrium ion abundances and consistently models the self-ionization of the shock  They found a spectrum that on average scales as $\alpha = 1.7$, between $10~$eV before a cutoff at $\sim 1~$keV.  Such a spectral index between $10^2-10^3$eV falls in between the effective scaling of our two considered $T_{\rm max}$ models.

\begin{figure}
\begin{center}
\resizebox{9.0cm}{!}{\includegraphics{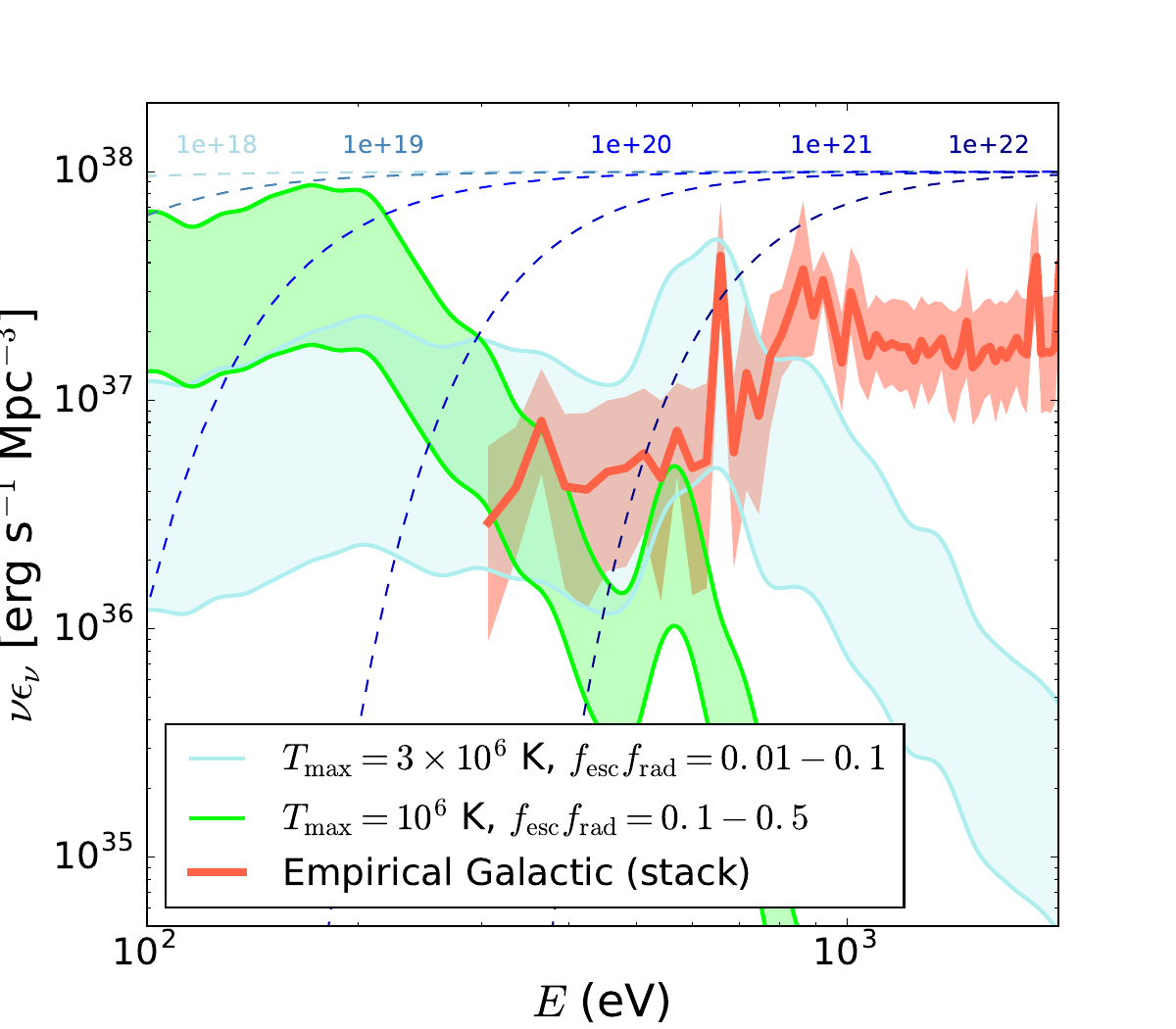}}\\
\end{center}
\caption{Estimates for the $z=0$ ISM and CGM emissivities using the models discussed in \S~\ref{sec:ISM}.  
 The turquoise band is our fiducial, theoretically-calculated ISM emissivity that uses equation~(\ref{eq:ISM_SNe}) with  $T_{\rm max} = 3\times 10^{6}$~K and $f_{\rm esc} f_{\rm rad} =0.01-0.1$. The green band is our CGM/ISM emissivity that also uses Equation~\ref{eq:ISM_SNe}, but with $T_{\rm max} = 10^{6}~$K and $f_{\rm esc} f_{\rm rad}=0.1-0.5$.  The red curve represents the stack of four star-forming galaxies (the empirical model in \S~\ref{sec:XRB}). The various blue dashed curves represent the amount of \HI\ + \HeI\ absorption for hydrogen column densities increasing from $10^{18}$cm$^{-2}$ to $10^{22}$cm$^{-2}$.
\label{fig:ISM_eps} }
\end{figure}

Our normalization of the ISM and CGM emissivities uses that the radiated energy is likely a significant fraction of the total feedback energy.  We consider supernova feedback, although stellar winds can be comparable energetically  \citep{leitherer99}.  The bolometric luminosity per unit SFR -- what sets the emissivity normalization in our models (eqn.~\ref{eqn:enu}) -- is
\begin{equation}
\label{eq:ISM_SNe}
[L_{\rm bol}/{\rm SFR}] = f_{\rm esc} f_{\rm rad}  f_{\rm SN} E_{\rm SN} ,
\end{equation}
where $f_{\rm SN}$ is the approximate fraction of stars expected to go supernova that we take to be $10^{-2}$, $E_{\rm SN}$ is the approximate energy output per supernova that we take to be $10^{51}$erg~s$^{-1}$, $f_{\rm rad}$ is the fraction of the feedback energy that is radiated, and $f_{\rm esc}$ is the fraction that escapes the galaxy. Calculations find that $f_{\rm rad} \approx 0.9$ in the local ISM surrounding supernovae if they are unclustered \citep{1998ApJ...500...95T}, and that for clustered supernovae this fraction can go down to $f_{\rm rad} \approx 0.5$ \citep{2014MNRAS.443.3463S}.  
  Much of the energy which is not radiated in the ISM (or, if radiated, that is re-absorbed) is likely radiated in the CGM where $f_{\rm esc} \approx 1$:  \citet{mcquinn17} argued that the large OVI columns require $f_{\rm rad} > 0.1$ in the CGM, and large $f_{\rm rad}$ appear to be born out in numerical simulations of the CGM \citep{fielding16, 2013MNRAS.430.2688V}.  Observations further suggest that the characteristic scale for the CGM emission extends many tens of kiloparsecs outside of the galaxy \citep{werk14}, and so this diffuse emission would likely go undetected in soft X-ray images even if the gas reached sufficiently high temperatures to be seen in the soft X-ray.   We note that this energetics-motivated model for CGM emission is similar to that of \citet[][except they assume emission at $T_{\rm vir}$]{miniati04}, as are our conclusions for the potential impact on $J_\nu$.

In our model of the CGM emission (though it can also be colder, less obscured ISM emission), we take the possible range for escaping emission to be $f_{\rm esc} f_{\rm rad}=0.1-0.5$ and $T_{\rm max} = 10^6$K. This `CGM/ISM' model is shown by the green band in Figure~\ref{fig:ISM_eps}.  This spectrum peaks at hundreds of eV, and falls off below the X-ray measurements at $>300$eV.  However, we again note that falling below the X-ray measurements is not required as this emission could be too diffuse to detect.

For hotter ISM gas with $T_{\rm max}\approx 3\times10^6~$K, we tune the normalization to meet the observations, finding $f_{\rm esc} f_{\rm rad} = 0.01-1$. This `ISM' model is shown by the turquoise band in Figure~\ref{fig:ISM_eps}.  Much of the ISM radiation is likely absorbed in the galaxy to produce such low $f_{\rm esc} f_{\rm rad}$: the blue dotted lines represent the \HI\ $+$ \HeI\ transmissions for \HI\ column densities ranging from $10^{18}$cm$^{-2}$ to $10^{22}$cm$^{-2}$.  Large \HI\ columns of $\sim10^{22}$cm$^{-2}$ are necessary to obscure the X-ray emission.  Additionally, lower columns would obscure ISM models with smaller $T_{\rm max}$.  

For both the ISM and CGM models, we assume the same range of $f_{\rm esc} f_{\rm rad} $ holds with redshift to compute $J_\nu$ in \S~\ref{sec:jnu}.  As the properties of galaxies change, and it is thought that the escape of ionizing photons goes up with redshift, this might underestimate temporal trends, especially for $J_\nu$ at higher redshifts.

\subsection{Virialized hot halo gas}
\label{sec:ff}
This section considers the emission from virialized gas in massive halos.  We distinguish this scenario from the diffuse CGM gas previously considered in that, as cooling times get longer in larger systems, a roughly single temperature hydrostatic virialized region forms, whereas the CGM emission owes to cooling gas and likely a range of temperatures.  A second distinctions is that the CGM emission is sourced by halos where most star formation occurs in the Universe, $\sim 10^{12}\Msun$ halos, whereas we will see that hot halo emission owes to systems at least an order of magnitude larger (and is powered by virialization and quasar feedback). 

The hot gas scenario should be thought of as being essentially the same as warm hot intergalactic medium emission that has been discussed extensively by prior studies as a source of the soft X-ray background \citep{1999ApJ...514....1C, 2001ApJ...552..473D, 2001ApJ...557...67C}. While some of the diffuse emission likely comes from filamentary gas, most is likely to come from gas associated with halos as modeled in \citet{1999ApJ...510L...1P}.  There is 25\% (15\%) of dark matter in $>10^{13} M_\odot$ halos at $z=0 ~(0.5)$, and 10\% (5\%) of dark matter in $>10^{14} M_\odot$ halos.  Filamentary gas is too diffuse (and, possibly, too unenriched) to compete with virialized regions even though it constitutes a somewhat larger mass fraction.  The amplitude of this emission is more sensitive to how the associated gas is distributed around halos, which we discuss below.

Group and cluster virialized gas emits in the EUV and soft X-ray of massive halos via free-free and atomic line emissions, with the latter becoming more prominent in lower temperature halos.  The X-ray emission from virialized gas that has been measured for $\gtrsim 10^{13}M_\odot$ halos can be directly observed \citep{jeltema08,holden02,markevitch98,helsdon00,mulchaey03,dellaceca00,arnaud99,borgani01,pratt09}, and we design our models to latch onto these observations.  In particular, we use the models of \citet{sharma12} based on timescale considerations, which are able to fit the observed X-ray luminosity -- temperature relation ($L_X$-$T_X$).  In particular, \citet{mccourt12} and \citet{sharma12} find (using simulations of a gravitationally stratified atmosphere in which cooling is initially in balance with heating) that in order for a halo to not be locally thermally unstable, leading to condensation and driving star formation rates larger than observed, the cooling time must be at least ten times the dynamical time.  This bound on the density of gas in halos has been confirmed in simulations with more realistic feedback prescriptions \citep{gaspari}.  It also is consistent with observations \citep{sharma12, 2015ApJ...799L...1V, 2017arXiv170802189V}.

 We use the \citet{sharma12} fiducial model which uses a cooling time to dynamical time ratio of 10 (an upper limit on the core density and therefore on $L_X$) as an upper limit to our calculation and their lowest luminosity model with a cooling time to dynamical time ratio of 100 as a lower limit.  These models, which produce $L_X^{\rm Sharma}$, generously bracket the spread in $L_X$-$T_X$ values seen in observations (see Figure 3 of \citealt{sharma12}).  Indeed, a more detailed analysis likely could shave off factors of two in the uncertainty at both sides.  Both of these models use an asymptotic gas density slope of -2.25 to solve a differential equation for the gas density profile (see \citealt{sharma12}).  We identify $T_X$ with $T_{\rm vir}(M_{\rm halo})$, the virial temperature of a halo of mass $M_{\rm halo}$, calculated assuming an isothermal sphere \citep{barkanaloebreview}.\footnote{We note that the Sharma models are computed for just free-free cooling, which could lead to differences at the lowest temperatures we consider where atomic cooling becomes important.  However, the CGM models become more applicable in this limit.}
 
 For the spectral form of this emission, $\ell(\nu) \equiv L_\nu/\int d\nu L_\nu$, we use a composite {\sc Cloudy} spectrum that is made up of a series of {\sc Cloudy} ``coronal equilibrium" models with several discrete temperatures.  These {\sc Cloudy} models assume collisional ionization equilibrium \citep{ferland13} and a hydrogen density of $10^{-4}$ cm$^{-3}$ -- the virial density of halos in spherical collapse at $z=0.2$. We use Solar abundance ratios of \citet{grevesse10} with an overall metallicity of $Z=0.3 \,Z_\odot$, in accord with the typical metallicity in intracluster and circumgalactic media measurements \citep{prochaska17,mantz17}. Changing to $Z=0$ (or, to a lesser extent, turning on an ionizing background or decreasing the density we assume) results in the emission lines disappearing in the spectra, but do not significantly change our results.  
 
The temperatures of these models are computed over a large range of virial temperatures. The models in the composite spectrum are then weighted in accordance to the Sheth-Tormen halo mass function \citep{sheth99}, $\frac{dn}{dM}$, at a given redshift such that
\begin{equation}
\epsilon_{\nu}= \int^{\infty}_{M_{\rm min}} dM \frac{dn}{dM} L^M (\nu),
\end{equation}
where $L^M (\nu)$ is the frequency-dependent luminosity of the halo gas of a given halo mass, set equal to $L_X^{\rm Sharma}\times \ell^{\rm cloudy}(\nu)$. We set $M_{\rm min} = 10^{13} \Msun$ here, but smaller values would not change our results for $\epsilon_\nu$ and $J_\nu$  (as discussed below) at $z=0$.

\begin{figure}
\begin{center}
\resizebox{9.0cm}{!}{\includegraphics{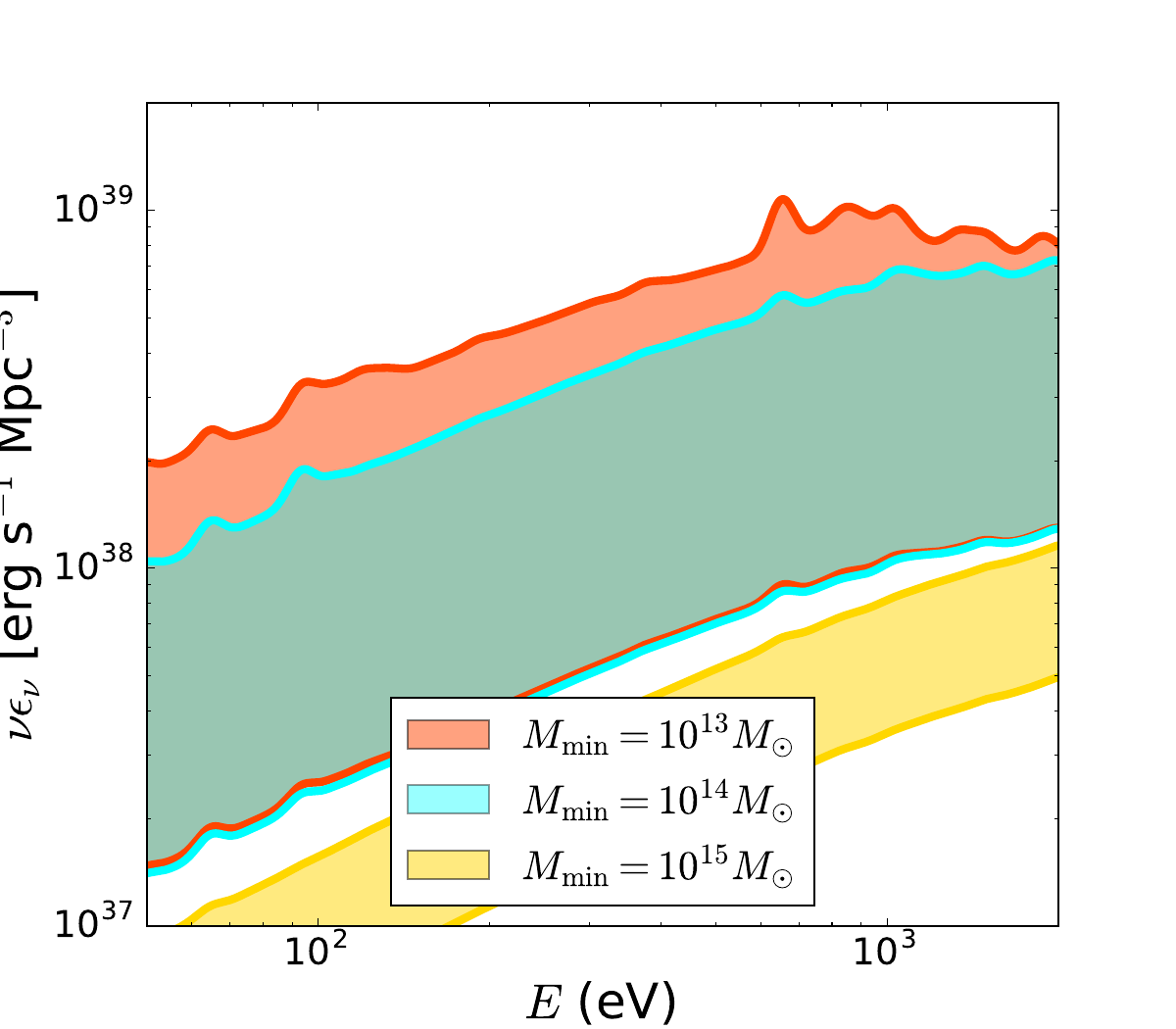}}\\
\end{center}
\caption{Specific emissivity of $z=0$ virialized halo gas following the calculation described in \S~\ref{sec:ff}. 
 The lower (upper) limit uses the models in \citet{sharma12} to relate the halo virial temperature and luminosity, assuming a cooling timescale to dynamical timescale ratio of 100 (10).  These values are chosen to bracket  observations. The red band is our fiducial model and represents a minimum halo mass of $10^{13} M_{\odot}$. The cyan and gold bands correspond to minimum masses of $10^{14} M_{\odot}$ and $10^{15} M_{\odot}$ respectively. }
\label{fig:ff_eps}
\end{figure}

Our fiducial model, shown as the red band in Figure~\ref{fig:ff_eps}, is this calculation for the $z=0$ emissivity of halo gas, with the band bracketing the two $T_X-L_X$ models of \citet{sharma12} described above. The cyan band and gold band show this calculation above higher minimum masses as discussed in the following paragraph. If we treat the observed spread as an indication of measurement error (or, e.g., biases owing to missing the most diffuse components), then the band indicates the level of uncertainty.  Interestingly, these estimates for virialized gas emission exceed our lower bounds for quasar emission between several tens and a few hundred eV. In contrast to the $z=0$ emissivity, the observed background comes from a range of redshifts, and when this is considered we find in Section~\ref{sec:jnu} that quasars are still likely the dominant source due to the declining abundance of massive halos at increasing redshift.  

Most of the $z=0$ emission from virialized gas comes from small clusters.  Figure~\ref{fig:ff_eps} also shows  calculations that only allow for emission above halos with masses of $10^{14} M_{\odot}$ (cyan band) and $10^{15} M_{\odot}$ (gold band), respectively. We find that the EUV and soft X-ray emission from this gas is dominated by $M\gtrsim 10^{13}\Msun$ at $z=0$.\footnote{Because the virial scale is an arc-minute for a $M = 10^{13}\Msun$ halo at $z=0.5$, these sources of emission would not be included in X-ray point source catalogues.  Nearly every sightline on the sky intersects a $10^{13}M_{\odot}$ halo \citep{mcq14}. }  

Lower mass halos cool very efficiently and so have lower gas densities.  In the \citet{sharma12} models, the gas in the most massive halos essentially traces the dark matter, but becomes cored at lower masses with a large fraction does not fitting within in the virial radius for $M\lesssim 10^{13}\Msun$.  We caution that at $M\lesssim 10^{13}\Msun$, representing the cooling emission with a virialized atmosphere may be fraught and the CGM calculations described previously likely hold more bearing.

\section{the contribution of different sources to $J_\nu$}
\label{sec:jnu}
\begin{figure*}
\begin{center}
\resizebox{19.5cm}{!}{\includegraphics{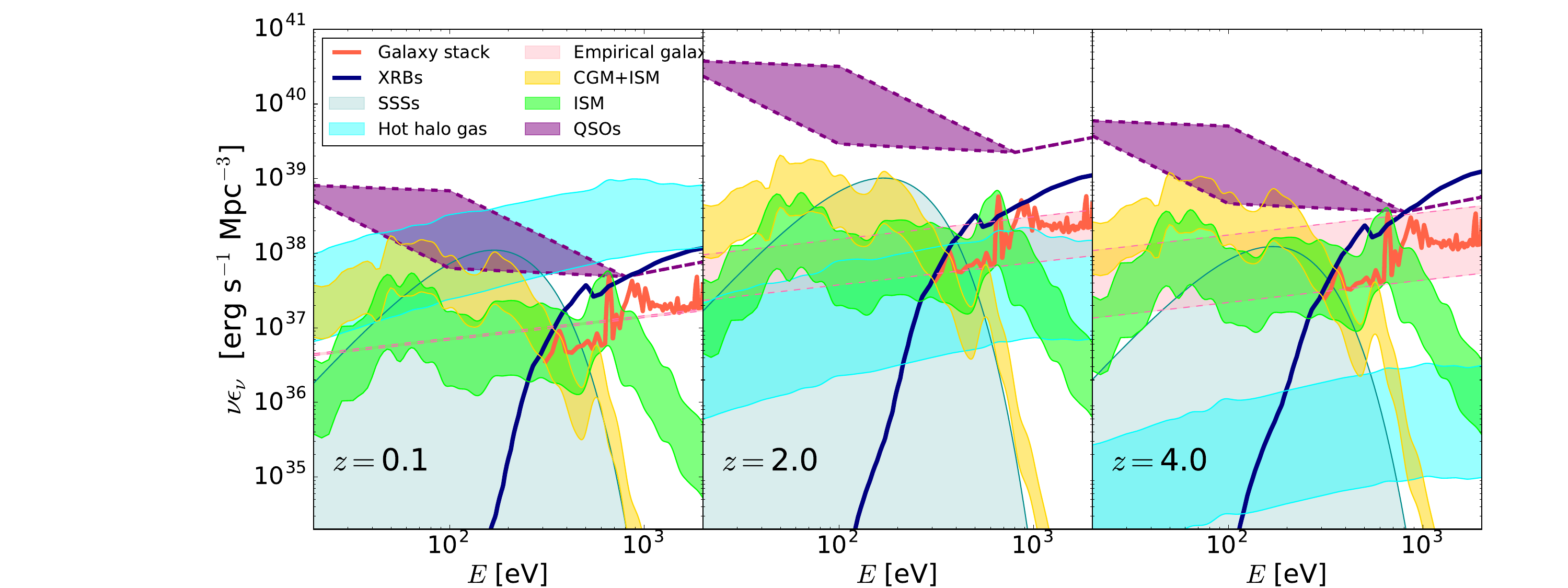}}\\
\end{center}
\caption{Estimated specific emissivities of the sources using the models described in Section~\ref{sec:sources}.  The purple shaded region shows quasar emissivity estimates from extrapolating into the EUV/soft X-ray with the range of mean power-law slopes that are estimated. The ISM/CGM are represented in the green and yellow bands, with $T_{\rm max}=3\times 10^{6}$ K and $T_{\rm max}=1\times10^{6}$ respectively, with their spectra set to a resolution of 15 to smooth out lines. The cyan band represents the virialized hot halo gas described in Section~\ref{sec:ff}.  The dark blue SED represents the \citet{fragos13b} XRB model.  The red spectrum represents the empirical XRB model from stacked spectra, and the light pink band the power-law extrapolation of galactic soft X-ray emission (whose width increases with redshift according to the allowed range of evolution in XRB emissions), and the teal shaded area represents our upper bound on the emissivity of super soft X-ray sources with an effective temperature of $5\times 10^5$ K}.
\label{fig:nu_e_nu}
\end{figure*}

The summary of the emissivity calculations described in Section~\ref{sec:sources} is shown in Figure~\ref{fig:nu_e_nu}. We plot each of the sources' specific emissivities at $z = 0.1$, $2.0$, and $4.0$. 
We use the quasar calculations described in Section~\ref{sec:QSO}, the XRB calculation of Section~\ref{sec:XRB}, the SSS calculation of Section~\ref{ss:SSS}, the ISM and CGM calculations as described in Section~\ref{sec:ISM}, the empirical galactic model with a power-law SED discussed in Section~\ref{sec:XRB}, and virialized halo gas calculation as described in Section~\ref{sec:ff}. The band encapsulating each model reflects the range of possible values that we motivated.

Both the peak of the quasar emissivity and cosmic star formation rate (to which the XRB and CGM emissivity are tied) occur at $z=2-3$, prompting a similar peak in each sources' emissivity near these redshifts.  In contrast, the amount of free-free emission from virialized halo gas drops off rapidly with increasing redshift due to the abundance of $\gtrsim10^{13}M_{\odot}$ halos decreasing quickly with redshift.  While virialized gas likely is the most important source at most of the considered wavelengths at $z=0$, quasars become the most important at $z=2-4$.  At higher redshifts than shown, models suggest that XRBs and perhaps hot ISM gas become the dominant source at $100-1000\;$eV \citep{2011A&A...528A.149M, 2014MNRAS.443..678P, 2017ApJ...840...39M}.

The solution to the cosmological radiative transfer equation gives the background angular-averaged specific intensity at a particular energy as observed at $z_0$,
\begin{equation}
J_{\nu}(z_0)=\frac{c}{4\pi}\int_{z_0}^{\infty}dz\left|\frac{dt}{dz}\right|\frac{(1+z_0)^3}{(1+z)^3}\epsilon_{\nu}(z)e^{-\bar{\tau}(z_0,z,\nu)},
\end{equation}
where $\epsilon_\nu$ is the proper volume emissivity -- what we modeled in Section~\ref{sec:sources} for various sources  --, $\left| dt/dz\right|=\left[H(z)(1+z)\right]^{-1}$, and $\bar{\tau}$ is the effective optical depth between $z_0$ and $z$ in a nonuniform IGM, and given by 
\begin{equation}
\bar{\tau}(E_0,z_0,z)= \int_{z_0}^{z}dz'\int_0^{\infty} dN_{\rm HI} f(N_{\rm HI},z')\left(1-e^{-\tau_\nu(N_{\rm HI}))}\right),
\end{equation}
for randomly distributed absorbers, where $f(N_{\rm HI},z')$ is the \HI\ column density distribution and $\tau_\nu(N_{\rm HI})$ is the optical depth at a particular energy through an absorber,
\begin{equation}
\tau_\nu = N_{\rm HI}\sigma_{\rm HI}(\nu) + N_{\rm HeII}\sigma_{\rm HeII}(\nu),
\end{equation}
where $\sigma_X$ is the photoionization cross section of species $X$, and the \HeII\ column of a system $N_{\rm HeII}$ must be modeled in terms of $N_{\rm HI}$.  We use Equation~A10 in \citet{fardal98} for the relationship between $N_{\rm HeII}$, $N_{\rm HI}$, $\Gamma_{\rm HI}$ for a single temperature and density slab.  We assume $10^4$K and a density set by the model in \citet{schaye01} in which absorbers column and density are related by assuming the size of absorbers is the Jeans length and photoionization equilibrium.  The \citet{schaye01} model agrees well with simulations \citep{mcquinn11, altay11}.  Our calculations neglect \HeI\ continuum absorption and resonant lines, which contribute relatively small features in $J_\nu$ compared the the \HI\ and \HeII\ continuum absorption that we include.\footnote{
Because the mapping of $N_{\rm HI}$ to $N_{\rm HeII}$ depends on $J_\nu$, we must solve for $J_{\nu}$ iteratively. 
  Additionally, we use the rates for our fiducial quasar model when calculating the generally subdominant contribution to $J_\nu$ from other sources models. }

For $f(N_{\rm HI},z)$, our calculations use the piecewise-in-$N_{\rm HI}$ fits of  \citet{prochaska14}, which use a parametrization of the form $f(N_{\rm HI},z) = A N_{\rm HI}^{-\beta}(1+z)^{\gamma}$.  While the error bars over the range of columns probed are generally quoted at the tens of percent level, there are some factor of two discrepancies around $10^{16}$cm$^{-2}$ \citep{prochaska14}.  However, for the $z\sim 0$ background, the dilution from expansion and redshifting (which is perfectly captured) is more important than the continuum absorption of hydrogen.  At $\gg1$Ry, where we focus, this is true not just at $z\sim 0$ but also at higher redshifts.  Therefore, discrepancies in the column density distributions may indicate there is room for as much as a factor of two difference from mean free path effects, but then likely only at the lowest energies considered and moderate redshifts.  Even a factor of two is smaller than the possible amplitude ranges in all of our source models.  Thus, our modeling of absorption is not the dominant source of uncertainty.

\begin{figure}
\begin{center}
\resizebox{9.5cm}{!}{\includegraphics{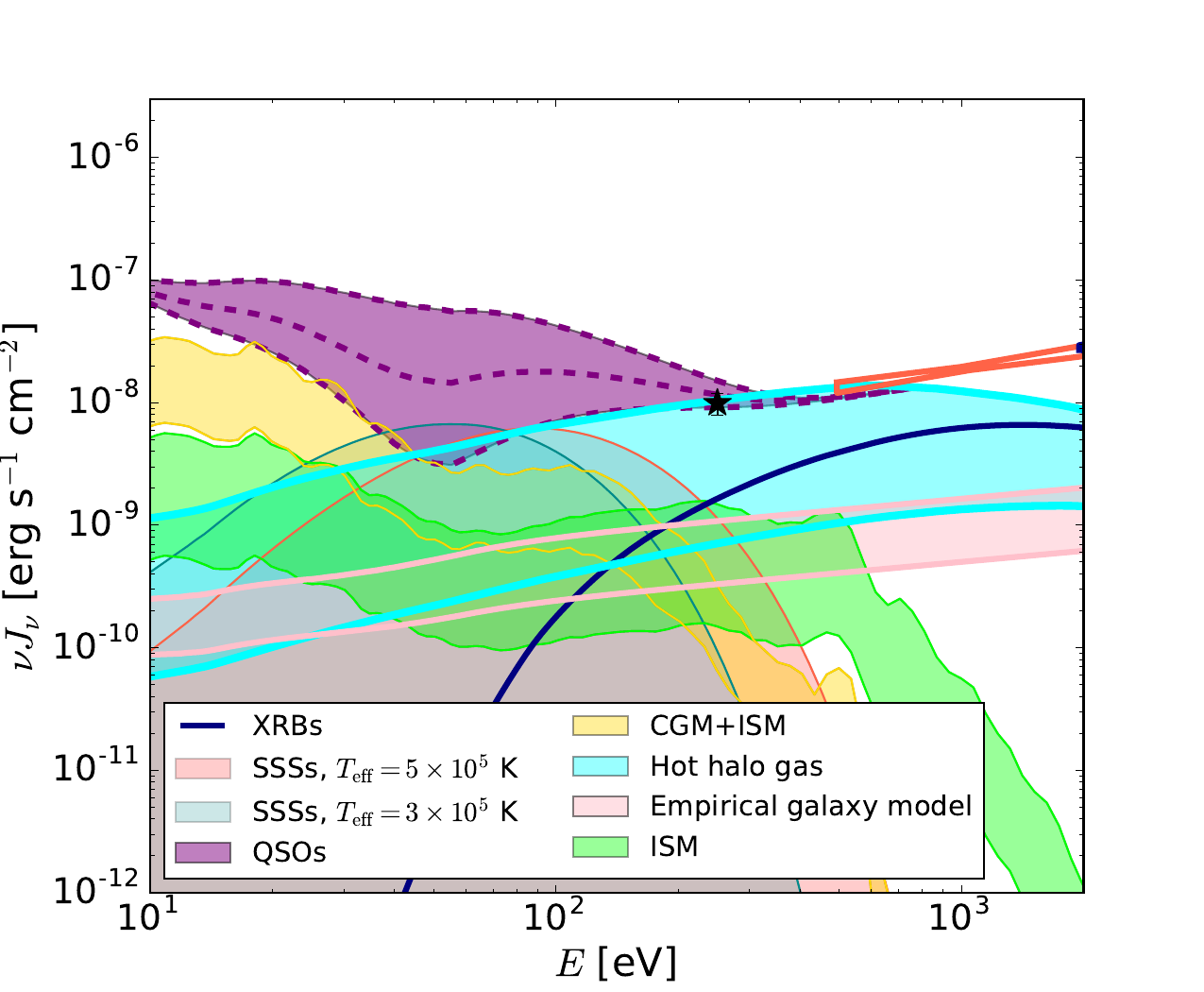}}\\
\end{center}
\caption{Angle-averaged $z=0$ specific intensity, $J_{\nu}$, of sources described in Section~\ref{sec:sources} and shown in Figure~\ref{fig:nu_e_nu}. The errorbar at $E=250$ eV is the soft X-ray background intensity {\it ROSAT} measurement by \citet{warwickroberts98}. The red bowtie is {\it ROSAT} PSPC observations from \citet{georgan96}. Despite observations suggesting that $\alpha_{\rm QSO}$ may be on the softer end, quasars likely dominate the background at energies $\lesssim$ 100 eV. Between one and several hundred electron volts, virialized hot halo gas may can become as important as quasars. Galactic sources such as XRBs, SSSs, the ISM, and especially the CGM, can contribute at the tens of percent level.}
\label{j_nu}
\end{figure}

Figure~\ref{j_nu} shows $J_{\nu}$ at $z=0$ for each of the sources shown in Figure~\ref{fig:nu_e_nu}. We also include two observational constraints from {\it ROSAT}: The constraint of \citet{warwickroberts98} using the lowest energies where the extragalactic background can be estimated is represented by the error bar at 250 eV, and the {\it ROSAT} PSPC data constraint of \citet{georgan96} is shown as the red bowtie starting at $\sim$500 eV. By construction, both of these observational bounds are reasonably matched by the full range of quasar emissivity models. 

Our calculations show that, even accounting for our estimated range of specific emissivities, quasars are the dominant contributor to the EUV/soft X-ray background at energies $\lesssim$100 eV.  If the value of $\alpha_{\rm QSO}$ is on the softer side as observations may suggest, then the emission of hot halo gas in massive halos could be the more important contributor to the EUV background at energies upwards of a couple hundred electron volts. Figure~\ref{j_nu} suggests that galactic sources may be of secondary importance to these two sources where $E\geq 100$ eV, but the ISM and CGM may be second only to quasars at lower energies.  Even though they are not the dominant sources of emission of the extragalactic background, they can dominate within the halos of galaxies (\S~\ref{sec:prox}) and their contribution.

Potential background sources can be isolated using unresolved X-ray background measurements.  In particular, the $1-2$~keV, resolved fraction of X-ray background is 80\% of the total in the Chandra deep fields (a total of 0.06 sq. deg.), and similar constraints have been found in larger fields \citep{2006ApJ...645...95H}.  Thus, only $\approx 20\%$ of the X-ray background owes to diffuse sources like virialized halo gas.  This bound suggests that the maximum virialized halo gas emission is a factor of $\sim 2$ below our maximum estimates.  However, at $1-2~$keV, the signal is dominated by rare $>10^{14} M_\odot$ halos (Fig.~\ref{fig:ff_eps}), which may not fall in the Chandra field of view.  A more detailed analysis is necessary to understand this constraint.

\begin{figure}
\begin{center}
\resizebox{9.5cm}{!}{\includegraphics{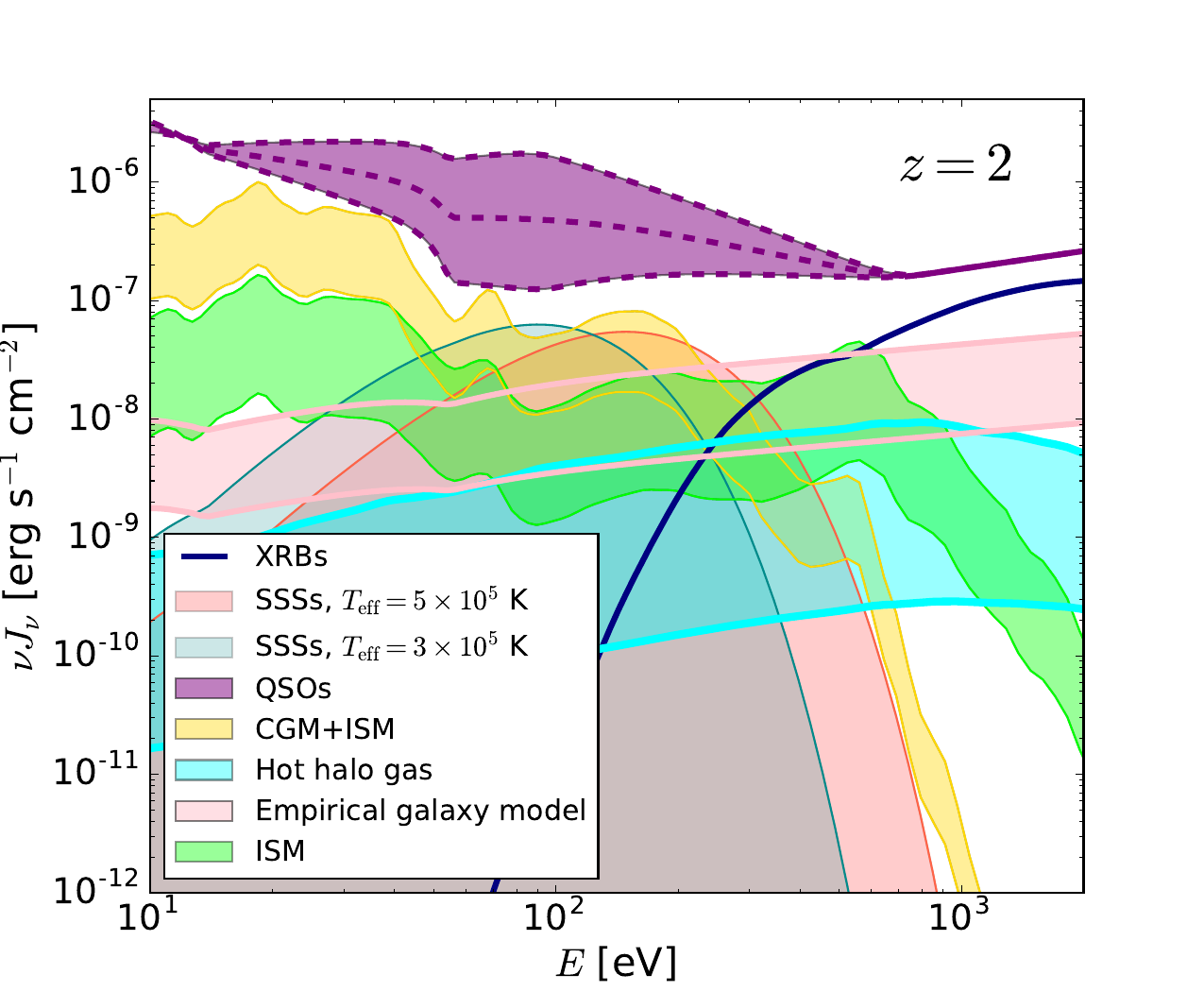}}\\
\end{center}
\caption{Angle-averaged $z=2$ specific intensity, $J_{\nu}$, of sources described in Section~\ref{sec:sources} and shown in Figure~\ref{fig:nu_e_nu}. At this redshift, quasars dominate the background at throughout the EUV and soft X-ray. Virialized hot halo gas is much less important than at $z=0$ due to the declining abundance of $M>10^{13}M_{\odot}$ halos.}
\label{j_nu_z2}
\end{figure}

Figure~\ref{j_nu_z2} shows $J_{\nu}$ for the same sources at $z=2$. Here quasars are undisputedly the dominating source of the EUV/soft X-ray background.  No other source is likely to produce enough $>54$eV photons to doubly ionize the \HeII\ if it is ionized at $z\sim 3$ as observations suggest \citep{2016ARA&A..54..313M}. The number of $M>10^{13}M_{\odot}$ halos steeply declines with increasing redshift, and thus the contribution of virialized hot halo gas at $z=2$ is negligible in comparison to the contribution of both quasars and galactic sources. The overall increase in the background intensity from $z=0$ to $z=2$ follows from the rise of both the cosmic star formation rate and quasar luminosity function.

\section{the effect of the extragalactic background on IGM and CGM absorption line inferences}
\label{sec:cgm}

\begin{figure}
\begin{center}
\resizebox{9.0cm}{!}{\includegraphics{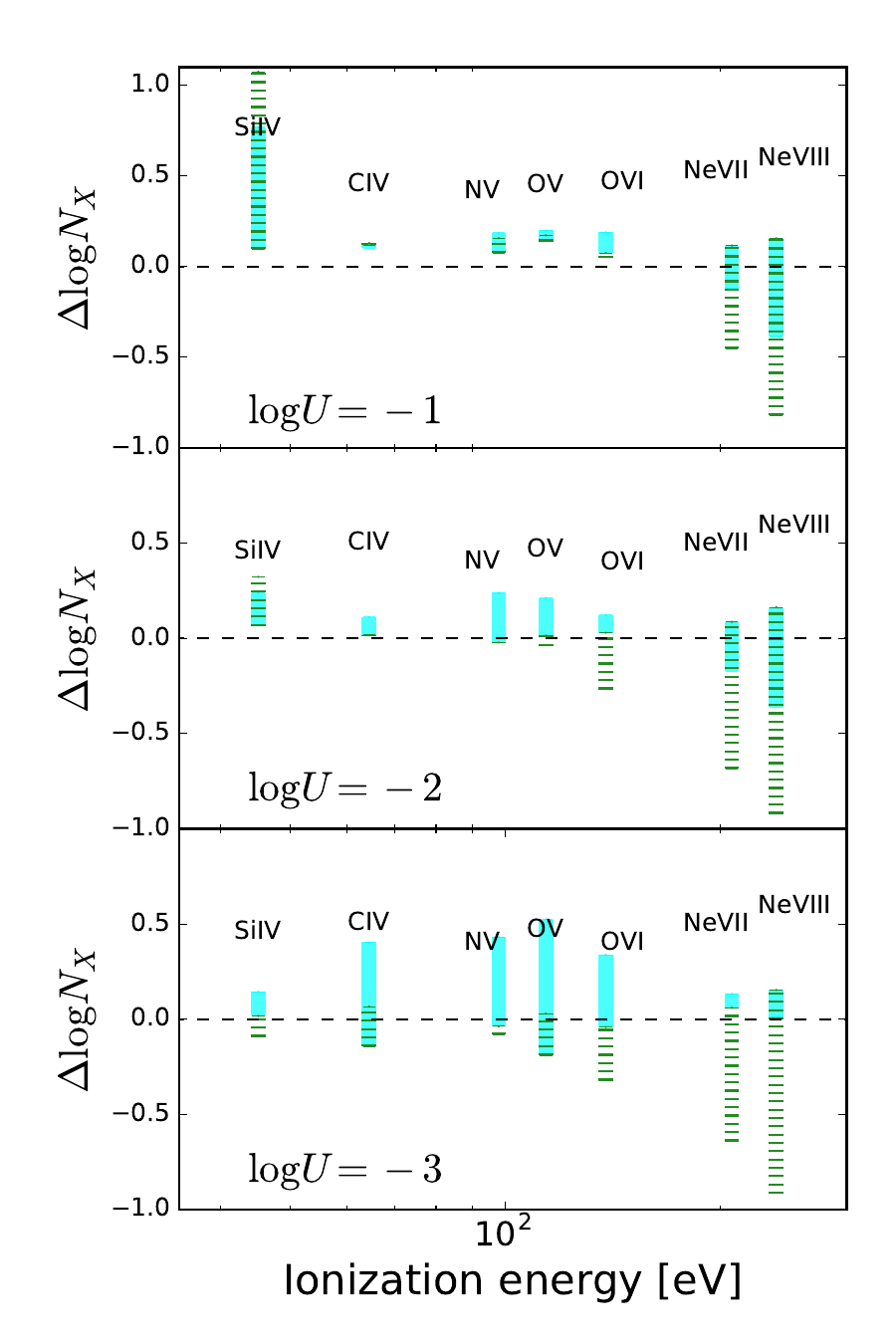}}\\
\end{center}
\caption{Error bars show estimates of column densities of several observationally relevant ions for the upper and lower limits of the $J_\nu$ range allowed, normalized to estimates using \citet{haardt12} background model and ordered by ionization energy of the least bound electron. The top panel shows an ionization parameter, $U$, of 10$^{-1}$, the middle 10$^{-2}$, and the bottom panel shows $U= 10^{-3}$, which approximately correspond to $n=10^{-5}$cm$^{-3}$, $10^{-4}$cm$^{-3}$ and $10^{-3}$cm$^{-3}$ for the \citet{haardt12} model.  The green dashed errors compare the differences at fixed $U$, and the cyan errors normalize the larger of our background models to a different $U$ in a manner that matches at the \HI\ photoionization rate of the smaller background model. \label{fig:N_ion}}
\end{figure}

Photoionization models are used to infer the density and metallicities of intergalactic and circumgalactic absorption systems \citep[e.g.][]{2016ARA&A..54..313M, tumlinson17}.  As ions have electronic binding energies in the EUV and soft X-ray, their abundances are shaped by these radiation backgrounds.  
  We have shown that the uncertainty in the spectrum of the background at these energies is considerable, suggesting commensurate uncertainties in the density and metallicity inferences.  In this section, we apply our models to quantify the sensitivity of the most observable ions to the assumed ionizing background.

Previous attempts to make metallicity and density inferences from ionic absorption measurements use a metagalactic ionizing background model, such as \citet{haardt12}.  When errors are quoted pertaining to the ionizing background choice, they are computed as the differences between historical models for these backgrounds, such as comparing how a calculation changes if it uses the \citet{haardt96} rather than the \citet{haardt12}.  The major differences between various historical models for the ionizing background is in improved constraints on IGM absorption and for the 1Ry emissivity of quasars, and not improved modeling of the spectrum of the sources.  In contrast, our models allow for us to quantify the principle source of uncertainty.

We use {\sc Cloudy} to compute the equilibrium column densities of several common metal ions in the UV, with our $J_{\nu}$ models as the input ionizing spectrum and assuming temperature equilibrium. We only consider the range of $J_\nu$ in our quasar models, using the lower and upper envelopes of the quasar intensity. We then compare the resulting columns to the columns with the \citet{haardt12} $J_\nu$.  These models implicitly assume that photoionization is dominant for all ionic species as the equilibrium temperature is $\sim 10^4$K.  (Collisional ionization is included, but since our calculations use the equilibrium temperature these effects are generally small.)   We vary the dimensionless ionization parameter, $U$, the ratio between the ionizing photon flux and the volume density of the gas.   We choose $\log_{10}U=-1, -2,$ and $-3$, which approximately bracket the mean of values of $U$ allowed by the data in \citet{werk14}. For \citet{haardt12}, these correspond to densities of $8\times 10^{-6}-8\times 10^{-4}$cm$^{-3}$, with larger $\log_{10} U$ corresponding to lower densities. However the precise density depends on the background model.

The dispersion from the range of models in select ions' column densities is shown in Figure~\ref{fig:N_ion}. Here we show the difference between the {\sc Cloudy} column densities computed with the lower and upper limits of our quasar $J_\nu$ band and those of our {\sc Cloudy} calculations using the \citet{haardt12} background model. (By only using the lower and upper parts of our band, we may underestimate range of expected columns if there are complex trends with $J_\nu$ in the ionic column.)  The green bars represent the range of $J_{\nu_{\rm QSO}}$ in relation to the \citet{haardt12} background model, $\log{N_{X, {\rm QSO}}}-\log{N_{X,{\rm HM12}}}$.  The reason the midpoint of our errors is not centered at zero, the \citet{haardt12} values, likely owes to our soft X-ray normalization being slightly lower than theirs.  The larger $\log{U}$ are more applicable for the higher ionization state metal lines (e.g. NeVIII), and the smaller $\log{U}$ are more applicable for the low ionization state metal lines (e.g. SiIV). We set $N_{\rm HI} =10^{15}$~cm$^{-2}$ and the metallicity to $Z = 0.3 Z_{\odot}$ in these calculations, but note that these are easily rescaled at fixed $U$ to higher column or more metal-rich systems. For the model with the \citet{haardt12} ionizing background, we find that with Log$U= -1$, the logarithmic column densities are \{10.0, 13.8, 14.0, 14.9, 15.1, 14.4, 14.0\} for [SiIV, CIV, NV, OV, OVI, NeVII, NeVIII], for Log $U= - 3$ they are \{12.1, 13.7, 13.1, 14.0, 13.3, 11.5, 10.4\}, and for Log $U= - 3$, the column densities are \{11.7, 11.9, 10.2, 11.2, 9.5, 6.7, 4.9\}. The dispersion in possible $J_\nu$ results in up to an order of magnitude in the uncertainty in the calculated column densities for the highest ions (Fig.~\ref{fig:N_ion}).

While normalizing to a fixed $U$ is standard practice in this field, normalizing to a fixed \HI\ photoionization rate would better encapsulate the uncertainty in the ionization correction, as this fixes the background to the most constrained location, as Ly$\alpha$ forest measurements nail down the \HI\ photoionization rate. While the green errors in Figure~\ref{fig:N_ion} compare the differences at fixed $U$, the cyan errors normalize the model that traces our upper envelope for the quasar $J_\nu$ to a different $U$ such that it matches the \HI\ photoionization rate of the lower envelope of $J_\nu$ model.  Because this normalization fixes the background at a lower energy (near $13.6$eV) compared to normalizing in $U$, this more physical normalization can sometimes result in more uncertainty, especially for the low ions.  Indeed, most of the low ions show $\approx 0.3$ dex variation.  When fixing the \HI\ photoionization rate, we find factors of $3-10$ errors in the ionization correction for all considered ions, which in turn reflects factor of $\sim3-10$ uncertainties in density and metallicity inferences from IGM and CGM absorbers.

\section{Proximity effects}
\label{sec:prox}

The uncertainties in derived metal ion column densities in the previous section neglect the possibility that the ionization state of circumgalactic gas could be affected by the ionizing flux of the host galaxy. Within some radius, local galactic sources will dominate the cosmic EUV/soft X-ray background in ionizing photon flux.  The low to mid ionization states of metals tend to arise in $10^4$K gas, with their ionization states set predominantly by the photoionizing radiation background.  Thus, knowledge of the UV background is critical for estimating the densities and metallicities of the clouds that these lower ions trace.  For example, the inferred densities for these clouds using standard ionizing background models is an order of magnitude smaller than predicted by pressure equilibrium with the virialized phase \citep{werk14, werk16, mcquinn17}. Modifications to the UV background have been hypothesized as a potential solution, likely requiring a $1-2$ order of magnitude enhancement from unknown sources at $\sim 100$eV \citep{werk16}.  In addition, because highly-ionized metals are major coolants of gas in the CGM, order unity changes in their ionization fractions from local emissions  translate to order unity changes in their cooling rates when the major coolants are predominantly photoionized, potentially impacting the rate at which galaxies are fed gas.

We compare the angular-averaged intensity from local galactic sources ( i.e. ISM, XRBs, and SSSs) to that from the global background to determine the ``proximity radius:" the distance from a galaxy where local and background sources contribute equally to the radiation.  The proximity radius for a galaxy is given by
\begin{equation}
\label{eq:r_prox}
r_{\nu, {\rm prox}} = 103~ {\rm kpc} ~\left[\frac{ ( [\nu L_{\nu}]/[10^{39} {\rm erg~s^{-1}}])}{ ( [\nu J_{\nu}]/[10^{-8} {\rm erg~s^{-1}cm^{2}]}) }\right]^{1/2}
\end{equation}
where  $L_{\nu}$ is the local galactic specific luminosity.  We have evaluated the equation at a $J_{\nu}$ characteristic of our ionizing background calculations, and for an $L_{\nu}$ at the maximum of what we find in our galactic source models.  Thus, we conclude that the proximity region is unlikely to extend beyond $\sim 100~$kpc, a smaller extent then inferred for many absorbers in the CGM and especially the \OVI\ \citep{tumlinson11, prochaska11, johnson15, werk16}.\footnote{See the appendix in \citet{mcquinn17} for generic arguments that the proximity region must have extent less than $\lesssim 100~kpc$.} We note that this expression treats sources as if they are residing at $r<r_{\nu, \rm prox}$; this will generally \emph{overestimate} the proximity radius in the case of emission extending beyond this radius (which may happen in the CGM). 

\begin{figure}
\begin{center}
\resizebox{9.0cm}{!}{\includegraphics{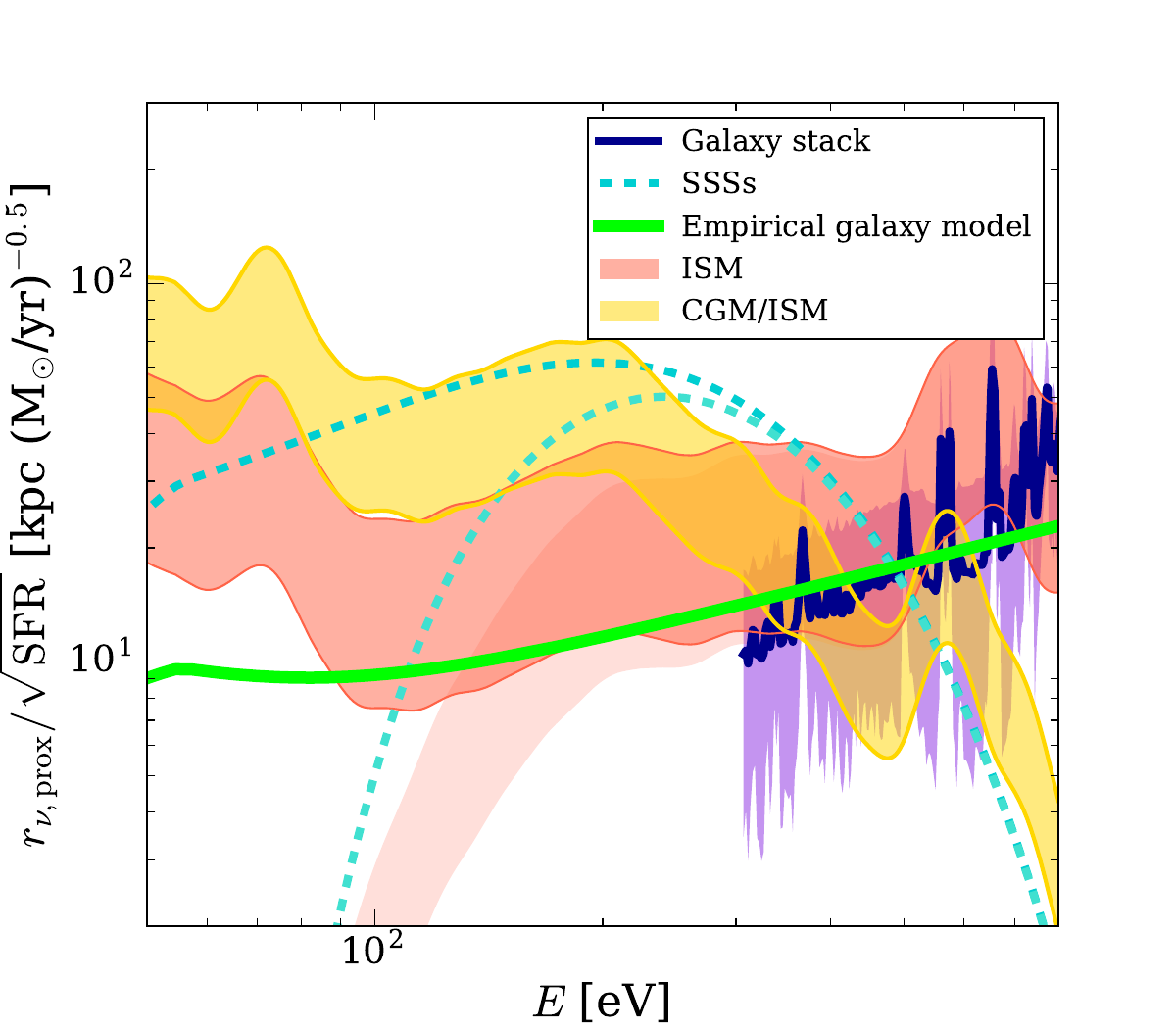}}\\
\end{center}
\caption{ The proximity region radius for a galaxy with a SFR of $1$ M$_{\odot}$ yr$^{-1}$ at $z=0.2$ for our galactic and circumgalactic source models, defined as the radius where the background $J_\nu$ is equal to the local contribution from galactic EUV and X-ray sources.  The fainter ISM band and the lower SSS curve represent the these models attenuated by a column of $10^{20}$ cm$^{-2}$. Because the luminosity of SSSs is expected to peak at stellar ages of about 1Gyr, the SSS models show the expected luminosity at $z=0.2$ from a galaxy with a SFR of $1.27$ M$_{\odot}$ yr$^{-1}$ at $z=0.3$ (1Gyr prior to $z=0.2$). This assumes that this toy galaxy's star formation history follows the shape of the cosmic SFR density.
\label{fig:r_prox}}
\end{figure}

Figure~\ref{fig:r_prox} shows our full calculations for the proximity radii in our models for $z=0.2$, a redshift chosen to match COS-Halos L* galaxy CGM survey \citep{tumlinson13}. 
 We show the galactic emission models described in Sections~\ref{sec:XRB},~\ref{ss:SSS}, and~\ref{sec:ISM} for the specific luminosities, $L_{\nu, \rm local}/$SFR.  For the extragalactic background, we use our fiducial quasar model with a power-law index of $\alpha = 1.7$.   In our source models, the proximity radius for a star-forming galaxy with a SFR of $\sim1{\rm M}_{\odot}$~yr$^{-1}$ varies between $10$ and $100~$kpc.  Taking the median of our bands or the observed X-ray luminosities would suggest $\sim 10-30$ kpc, a small fraction of the total CGM, but scaling as (SFR)$^{1/2}$. Thus, proximity radiation changes the nature of cooling gas that resides fairly close to galaxies but does not affect the bulk of the halo gas with extent $\sim 100~$kpc.  Thus, local sources are unlikely to be dominant for the vast majority of absorbers in many CGM samples, such as in \citet{tumlinson11} and \citet{stocke13}, whose absorbers mainly reside at $30-150~$kpc. That being said, assuming the radiation escapes, proximity emission can substantially affect gas ionization near galaxies and, therefore, it can
change the CGM gas cooling rate for the ions that are
sensitive to the radiation with energies in the range
shown in Figure~\ref{fig:r_prox} .

This conclusion is consistent with the geometric arguments of \citet{mcquinn17} that showed that $r_{\rm prox} \lesssim 100~$kpc, assuming that the proximity region sources are also the dominant source of the background.  Our calculations add the likely-dominant background from quasar emissions, which reduces this radius further beyond the \citet{mcquinn17} estimates.

\section{Conclusions}

We have modeled the sources of the extragalactic background in the EUV and soft X-ray.   In addition to the contribution from quasars, we included emissions that have so far been neglected in widely-used background models: X-ray binaries, the warm-hot interstellar and circumgalactic gas of star-forming galaxies, and virialized halo gas from groups and clusters.  Our models are calibrated against the latest observational measurements to bracket the possible range of contribution from each of these sources.  Note that these estimated ranges should be thought of as rough guidelines rather than rigorous bounds.

In agreement with previous studies, we find that quasars are likely the most important contributor to the ionizing background\footnote{With the caveat that stellar emissions -- which we do not model here -- are likely to be important at $<50$eV, with many models suggesting that they contribute up to a the factor of $2$ level and are possibly dominant at $z\gtrsim 4$ \citep[e.g.,][]{haardt12}.}.  However, we also show that their contribution to the observationally elusive $20-500~$ eV band can vary by up to a factor of $10$ for plausible extrapolations into this band using power-laws derived in the far UV and soft X-ray.  Furthermore, the evidence that the quasar SED is a single power-law, as assumed in previous background studies, is weak.  

If quasars contribute at the lower envelope of our estimated $J_\nu$ range, we find that emissions from the galactic ISM/CGM, from virialized halo gas in groups/clusters, and possibly from super soft sources may contribute significantly to the background. 
\begin{itemize}
\item The ISM contribution originates from cooling gas within stellar wind bubbles and supernovae blastwaves.  While a large fraction of the ISM emission is absorbed by \HI\ within the galaxy, a significant amount of stellar feedback energy is likely dumped as lower temperature gas and further away from obscuring \HI\ in the ISM, venting into the CGM, where it is more easily radiated away.  If most of the energy in stellar feedback is converted into background radiation, we find that the combined ISM/CGM contribution is roughly equal to the lower bound on quasars in the energy range $20-50$ eV. 
\item  We also find that, at $z=0$, emissions from hot, virialized gas in groups/clusters can be as important as quasars at energies of $\sim 100-1000~$eV.  This contribution falls off in importance at higher redshifts owing to the decreasing abundance of groups and clusters, and to the increasing number density of quasars. 
\item  Super soft sources have are often invoked as a huge unknown contribution at $\sim 100~$eV, although there is little empirical evidence for such a population.  By deriving a stringent bound from luminosity predictions of SSSs from previous models, we show that unseen super soft sources could maximally contribute tens of percent to the background at $\sim 100~$eV.
\end{itemize}

Disentangling the physics of circumgalactic gas with absorption line spectra requires accurate modeling of the EUV and soft X-ray backgrounds.  Previous studies have estimated the effect of uncertainties in these backgrounds by comparing inferences obtained from different historical background models -- for example, the \citet{haardt96} and \citet{haardt12} models.  As an application of our work, we have used our range of $J_\nu$ models to quantify this uncertainty more rigorously.  We bracketed the effect of the ionizing background on the inferred column densities of ions commonly identified in UV absorption line studies of the CGM.    We showed that uncertainties in the average quasar spectrum alone are sizable, enough to introduce a few tenths of dex differences in the abundances of the most observationally important ions and up to $1$~dex in the inferred column densities for the highest observationally-relevant ionization states.  These differenes likely translate into comparable uncertainties in the densities and metallicities constrained by these ions.

The uncertainties in metal ionization corrections are even larger if local sources -- e.g. XRBs and warm-hot ISM/CGM gases-- contribute significantly.  We estimated the EUV/soft X-ray proximity radius of a star-forming galaxy -- the radius at which local emissions become equal to the extragalactic background.  We found that this proximity radius is between $r_{\rm prox}\approx 10~$kpc and $100$~kpc, assuming a star formation rate of 1~$M_{\odot}~$yr$^{-1}$, with $r_{\rm prox}\approx 100~$kpc for choices that maximize the potential galactic luminosity (and only at $E <100~$eV).  Thus, local emission is unlikely to be the dominant source of ionization for the absorbers in most extragalactic CGM samples.

Future observations and theoretical work has the potential to further constrain emission process that contribute sizably to the EUV and soft X-ray backgrounds.   With a factor of $2-3$ improved sensitivity, diffuse soft X-ray background measurements would reach most of our model space for hot halo emission.   In analogy to the $\gamma$-ray, where angular anisotropy analyses have constrained the blazar, millisecond pulsar, and galactic contributions \citep{2012PhRvD..85h3007A}, a future soft X-ray space telescope could measure the angular anisotropy to constrain source models:  Low redshift clusters and groups are rare and have large arc-minute extents, whereas (abundant) galactic sources will contribute a smoother signal in angle. Finally, future CGM observations and modeling has the potential to better constrain the density and thermal structure of this medium and, hence, its cooling emissions.

\bigskip

This work was supported by NSF award AST 1614439, by NASA through the Space Telescope Science Institute awards HST-AR-14307 and HST-AR-14575, and by the Alfred P. Sloan foundation. We thank the referee for useful comments and suggestions. PRUS thanks Nell Byler for helpful discussion and MM thanks the Institute for Advanced Study visiting faculty program and the John Bahcall fellowship for support.

\bibliography{euvxray}

\end{document}